\documentclass[a4paper,11pt]{article}

\pdfoutput=1

\usepackage{graphicx}
\usepackage{latexsym}
\usepackage{amsfonts}
\usepackage{amssymb}
\usepackage{amsmath,bm}
\usepackage{mathtools}
\usepackage{url}
\usepackage{color}
\usepackage{multirow,array}
\usepackage{float}
\usepackage{dcolumn} 
\usepackage{rotating}

\usepackage{jcappub} 

\usepackage{tikz}
\usetikzlibrary{positioning}
\usetikzlibrary{calc}
\usetikzlibrary{backgrounds}
\makeatletter
\newcommand\footnoteref[1]{\protected@xdef\@thefnmark{\ref{#1}}\@footnotemark}
\makeatother

\newcommand{\Gaia}{\textit{Gaia}}

\newcommand{\appropto}{\mathrel{\vcenter{
  \offinterlineskip\halign{\hfil$##$\cr
    \propto\cr\noalign{\kern2pt}\sim\cr\noalign{\kern-2pt}}}}}

\newcommand{\approximatelylessthan}{\mathrel{\vcenter{
  \offinterlineskip\halign{\hfil$##$\cr
    <\cr\noalign{\kern1pt}\sim\cr\noalign{\kern-1pt}}}}}

\title{Sweeping the Dust Away -- Correcting the Phase Space Density of the Milky Way with Unsupervised Machine Learning}

\author[a]{Eric Putney,}
\author[a]{David Shih,}
\author[b,a]{Sung Hak Lim,}
\author[a]{Matthew R.~Buckley}

\affiliation[a]{NHETC, Department of Physics and Astronomy,\\Rutgers, the State University of New Jersey,\\126 Frelinghuysen Road, Piscataway, NJ 08854, USA}
\affiliation[b]{Particle Theory and Cosmology Group, Center for Theoretical Physics of the Universe, \\Institute for Basic Science (IBS), \\55 Expo-ro, Yuseong-gu, Daejeon 34126, Republic of Korea}

\emailAdd{eputney@physics.rutgers.edu}
\emailAdd{shih@physics.rutgers.edu}
\emailAdd{sunghak.lim@ibs.re.kr}
\emailAdd{mbuckley@physics.rutgers.edu}

\abstract{
The Boltzmann equation relates the equilibrium phase space distribution of stars in the Milky Way to the Galaxy's gravitational potential. However, observations of stellar populations are biased by extinction from foreground dust, which complicates measurements of the potential in the disk and towards the Galactic center. Using the kinematics of Red Clump and Red Branch stars in \Gaia{} DR3, we use machine learning to simultaneously estimate both the unbiased stellar phase space density and the gravitational potential. The unbiased phase space density is obtained through a learned ``dust efficiency factor" -- an observational selection function that accounts for dust extinction. The potential and the dust efficiency are parameterized by fully connected neural networks and are completely data driven. We validate the dust efficiency using a recent three-dimensional dust map in this work, and examine the potential in a companion paper.
}

\arxivnumber{2412.14236}  

\usepackage{fancyhdr}
\fancypagestyle{titlepage}{
  \fancyhf{}
  \fancyhead[R]{\raggedleft CTPU-PTC-24-38}
}
\fancypagestyle{otherpage}{
    \fancyhf{}
    \fancyhead[R]{}
    \fancyhead[L]{}
    \fancyhead[C]{}
}

\begin{document}
\maketitle
\thispagestyle{titlepage}  
\thispagestyle{otherpage}
\flushbottom

\section{Introduction}\label{sec:introduction}
Dark matter is unequivocal evidence of physics beyond the Standard Model. Though the nature of dark matter is of fundamental interest to particle physics, it must be remembered that all lines of evidence for the existence of dark matter -- as well as nontrivial constraints on its particle interactions -- come from astrophysics and cosmology \cite{1980ApJ...238..471R,1939LicOB..19...41B,Salucci:2018hqu,2011ARA&A..49..409A,1933AcHPh...6..110Z,Planck:2018vyg,Clowe:2003tk}. Of particular relevance to this work, the density profile of the Milky Way's dark matter halo is an important component of studies of dark matter. For example, the local density of our dark matter halo sets the expected scattering rate with visible matter in dark matter direct detection experiments (see Ref.~\cite{lim2023mapping} for a compilation of recent measurements and Refs.~\cite{2014JPhG...41f3101R, 2017MNRAS.465...76M} for detailed discussions). Additionally, non-trivial modifications of the dark sector, such as self-interactions or a component of ``warm" dark matter, can alter the expected halo density profile towards the Galactic center (see Refs.~\cite{Buckley:2017ijx, 2000ApJ...534L.143B, 2000ApJ...543..514K, 2000ApJ...544L..87Y, 2001ApJ...547..574D, 2002ApJ...581..777C, 2011MNRAS.415.1125K, 2012MNRAS.423.3740V, 2013MNRAS.430...81R, 2013MNRAS.430..105P, 2013MNRAS.431L..20Z, 2018ApJ...853..109E}).

One approach to measuring the local distribution of dark matter within the Galaxy is to treat the stars as tracers of the underlying gravitational potential $\Phi$. The stellar phase space density (PSD) $f(\vec{x},\vec{v})$, which is a function of position $\vec{x}$ and velocity $\vec{v}$, obeys the collisionless Boltzmann Equation (CBE):
\begin{equation} \label{eq:CBE}
\frac{\partial f}{\partial t} + \vec{v} \cdot \vec{\nabla} f = \vec{\nabla}\Phi \cdot \frac{\partial f}{\partial \vec{v}}.
\end{equation}
Historically,  measurements of $f$ (and its derivatives) with sufficient accuracy to yield stable and tractable solutions to the CBE were not possible, due in large part to the high dimensionality (six) of the phase space. Instead, moments of the velocity are obtained from Eq.~\eqref{eq:CBE} under the assumption of equilibrium ($\partial f/\partial t = 0$). The first such moments result in the Jeans Equations. These equations, along with binning stellar kinematic data under assumptions of symmetry, have formed the basis of many of the measurements of the Milky Way potential \cite{2018MNRAS.478.1677S,2020A&A...643A..75S,2020MNRAS.494.6001N,2021ApJ...916..112N,2020MNRAS.495.4828G,2018A&A...615A..99H}. 

Machine learning algorithms, in particular normalizing flows (see Refs.~\cite{9089305, 2019arXiv191202762P}), allow 
modeling of $f$ (and its gradients) from sparser data than previously possible.  This opens a new avenue to solving the CBE for $\Phi$ directly, using flows trained on data to represent the previously-inaccessible PSD. Additionally, flow models of $f$ are flexible and do not require any symmetry assumptions, allowing for a data-driven measurement of $\Phi$ with minimal assumptions.

Solutions to the CBE, which are under-constrained at a single point in phase space $(\vec{x},\vec{v})$, rely on the velocity-independence of $\Phi(\vec{x})$. A key insight from previous works was that all three components of $\vec{a}=-\vec{\nabla}\Phi$ can be over-constrained by constructing a system of multiple copies of Eq.~\eqref{eq:CBE} evaluated at a single $\vec{x}$ and many $\vec{v}$.
From this over-constrained system of equations, a single solution for $\vec{a}$ is obtained by treating the mean-squared error of the equilibrium CBE as a loss term and finding the single acceleration at each $\vec{x}$ that minimizes this loss.
To achieve the best numerical precision for $f$ and its gradients, sampled $\vec{v}$ are drawn from regions of phase space where $f(\vec{x},\vec{v})\neq0$. Generative models like normalizing flows are well-suited for this task: factorizing $f$ into the stellar number density $n(\vec{x})$ and the conditional velocity distribution $p(\vec{v}|\vec{x})$ that are each modeled by individual flows, the flow modeling $p(\vec{v}|\vec{x})$ can sample multiple $\vec{v}$ likely to be measured at a given $\vec{x}$.

Taking the second derivative of $\Phi$ results in the mass density $\rho$ within the Galaxy via the Poisson Equation. This technique was proposed and applied to mock potentials and stellar distributions in Refs.~\cite{2023ApJ...942...26G,2021MNRAS.506.5721A, 10.1093/mnras/stac153}, and demonstrated in Ref.~\cite{2023MNRAS.521.5100B} using synthetic stars drawn from a fully cosmological $N$-body simulation with realistic errors. In Ref.~\cite{lim2023mapping}, the authors of this work applied this technique to \Gaia{}'s  $3^{\rm rd}$ Data Release (DR3) \cite{2021Gaia,2022arXiv220800211G}. In this measurement, normalizing flows were trained on the six-dimensional kinematics of nearby stars as measured by the \Gaia{} Space Observatory's \cite{2016Gaia} radial velocity spectrometer (RVS) to provide the estimates of $f(\vec{x},\vec{v})$. Using this technique, we presented a data-driven map of the dark matter density in select regions of the local Milky Way.
However, this study was limited by several factors:
\begin{enumerate}\label{list:limitations}

\item The observed PSD $f_{\rm obs}$ is suppressed by dust extinction, obscuring significant portions of the disk. Unbiased solutions to Eq.~\eqref{eq:CBE} using $f_{\rm obs}$ as as estimate for $f$ could only be computed in carefully selected dust-free regions of the sky.\label{listitem:limitation_1}

\item The acceleration $\vec{a}$ was computed pointwise across the analysis volume instead of obtaining a continuous function. Consequently, we could not enforce the curl-free condition $\nabla\times\vec a=0$, 
nor could we prioritize solutions with a positive-definite mass density $\rho>0$.\label{listitem:limitation_2}

\item Estimates of $\rho$ were noisy and needed to be smoothed over a kernel $K$ as part of the numeric differentiation of $\vec{a}$. This limited the spatial resolution of $\rho$ and required computationally expensive sampling.\label{listitem:limitation_3}

\end{enumerate}

\begin{figure}[t]
    \centering
    \includegraphics[width=0.95\columnwidth]{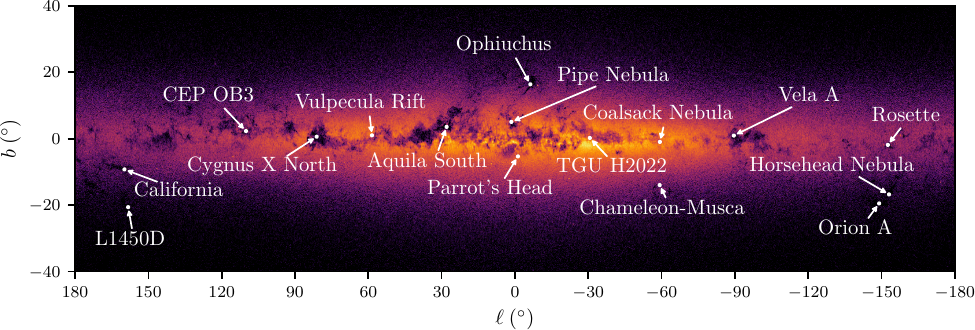}
    \caption{Projection of \Gaia{} DR3 RC/RGB stars within 4 kpc used for training the normalizing flows used in this analysis. The names and locations of prominent nearby dust clouds (dark nebulae (DNe) and molecular clouds (MoC)) are overlaid to contextualize the irregular and dark low-number density patches of the sky.
    The dark patch at Baade's window near the Parrot's Head Nebula $(\ell,b)\approx(1^\circ,-3.66^\circ)$ is an extremely crowded star field near the Galactic center. This crowding saturates \Gaia{}'s instruments, decreasing observational completeness in this region (Ref.~\cite{2021A&A...649A...5F}).
    }
    \label{fig:gaia_density_skymap}
\end{figure}

The first point -- dust extinction and completeness -- is a major obstacle for any analysis of the CBE: the sample of stars from which the PSD is constructed must be uniformly observed within the entire region of interest. Within the Galactic plane, and especially towards the Galactic center, the \Gaia{} DR3 dataset is incomplete due to the presence of dust, which both reddens and dims stars along the line of sight. This effect is clearly visible in Figure~\ref{fig:gaia_density_skymap}, where we show the observed number density  $n(\vec{x})$ as a function of location on the sky for bright stars
whose radial velocities were measured by \Gaia{}.
The visible dark patches are lines of sight along which stars have been dimmed by dust below the detection sensitivity of the \Gaia{} spectrometer; from this we can clearly see that the observed stellar density is reduced from a smooth equilibrium population along specific lines of sight.

This suppression of the observed PSD $f_{\rm obs}$ by dust makes it difficult to use Boltzmann techniques to measure the gravitational acceleration and potential (as well as the dark matter mass density) in the disk and towards the Galactic center. In Ref.~\cite{lim2023mapping}, we restricted ourselves to measurements along lines of sight either perpendicular to the disk or offset from it, in order to avoid the most significant areas of dust.

The dusty features of the Milky Way and their implications for the completeness of surveys such as \Gaia{} have been well-studied (Refs.~\cite{2023A&A...674A..31D, 2023A&A...669A..55C, 2020PASP..132g4501R, Gaia2mass2022}). However, using these dust maps to infer the true PSD of stars from the observed values is difficult. Inverting an extinction function to create a selection efficiency function requires knowledge of the concealed stellar population. Additionally, small errors in estimating this observational selection function or efficiency will result in large errors in the derivatives of the PSD. These errors then propagate in to the inferred values of the gravitational acceleration through the CBE.

In this work, we instead present a data-driven approach to learning the true stellar PSD from the observed stars by simultaneously learning the potential $\Phi(\vec{x})$ and a position-dependent selection function (or ``efficiency factor") $\epsilon(\vec{x})$ that encodes the suppression of the dust-corrected PSD ($f_{\rm corr}$) due to dust extinction:
\begin{equation}\label{eq:epsilon_def}
f_{\rm obs}(\vec{x},\vec{v}) = \epsilon(\vec{x}) f_{\rm corr}(\vec{x},\vec{v}).
\end{equation}
The efficiency $\epsilon(\vec{x})$ can be thought of as the integrated effect of the dimming due to dust applied to the particular population of stars along a line of sight. Up to an irrelevant normalization factor and the assumption that $\epsilon(\vec{x})$ is velocity independent (discussed in further detail in Appendix~\ref{app:eps_velo}), we expect $f_{\rm corr}$ to approximate the true distribution $f$ and better satisfy the equilibrium CBE than $f_{\rm obs}$. We rewrite Eq.~\eqref{eq:CBE} in terms of $f_{\rm obs}$ and $\epsilon(\vec{x})$:
\begin{equation}
\label{eq:CBE_eps}
\vec{v}\cdot\vec{\nabla} \ln f_{\rm obs} - \vec{v}\cdot\vec{\nabla} \ln \epsilon (\vec{x}) - \vec{\nabla}\Phi(\vec{x})\cdot \frac{\partial \ln f_{\rm obs}}{\partial \vec{v}} = 0.
\end{equation}
The introduction of $\epsilon(\vec{x})$ -- an additional position-dependent function in the CBE -- into Eq.~\eqref{eq:CBE_eps} increases the degrees of freedom to six and introduces a potential degeneracy between the solutions for $\epsilon$ and $\Phi$. Fortunately, this degeneracy is broken by the different couplings between the gradients of $\epsilon$ and $\Phi$ and velocity dependent quantities. $\vec{\nabla}\ln{\epsilon}$ couples directly to the velocity $\vec{v}$, whereas $\vec{\nabla}\Phi$ couples to the velocity gradient of the observed PSD $\partial \ln f_{\rm obs}/\partial \vec{v}$. Therefore, given $N_v\geq6$ unique $\vec{v}$ at a specified $\vec{x}$, we can solve the system of $N_v$ copies of Eq.~\eqref{eq:CBE_eps} for both $\epsilon$ and $\Phi$.

This novel approach to dust correction addresses point \hyperref[listitem:limitation_1]{1} above and extends our previous work to nearly the entire $4$~kpc ball around the Solar location.
To address points \hyperref[listitem:limitation_2]{2} and \hyperref[listitem:limitation_3]{3} above, we choose to parametrize both the log-efficiency $\ln\epsilon$ and the gravitational potential $\Phi$ with feed-forward neural networks (NN). This automatically guarantees the curl-free condition for $\vec{a}(\vec{x})$ and also allows us to include physical constraints such as positive-definite mass density. In addition, the use of a free-form model for $\Phi$ balances the computational efficiency of a global model for $\Phi$ (as opposed to the pointwise approach adopted in our previous work) without introducing biases that arise from restrictions to smaller spaces of functional forms. The idea of parameterizing $\Phi$ with a NN was first proposed in Ref.~\cite{2023ApJ...942...26G}, but this is the first work to introduce an efficiency function into the CBE and simultaneously learn both $\Phi$ and $\epsilon$.

The outline of our paper is as follows. In Section~\ref{sec:dataset}, we describe the dataset used to learn the observed six-dimensional PSD $f_{\rm obs}$ of the local Milky Way. In Section~\ref{sec:training}, we review and describe updates to the masked autoregressive flow (MAF) architecture and training techniques introduced in our prior work Ref.~\cite{lim2023mapping}, and then describe the training regimen for the $\Phi(\vec{x})$ and $\epsilon(\vec{x})$ neural networks. In Section~\ref{sec:results}, we compare our learned $\epsilon(\vec{x})$ to a state-of-the-art three-dimensional dust map, and then examine the dust-corrected phase space density. We reserve a detailed analysis of $\Phi(\vec{x})$ and downstream observables such as the three-dimensional acceleration and dark matter density fields for a companion paper, Ref.~\cite{dustpaperII}.

\section{Dataset}\label{sec:dataset}

\subsection{Coordinate System}
Our Cartesian coordinate system is a right-handed Galactocentric coordinate system, where the center of the Galaxy is at the origin. The Sun lies close to the $x$ axis at $x=8.122$~kpc and $y=0$~kpc. The Galactic midplane is defined as $z=0$, with the Sun being slightly off-set above the disk at $z=0.0208$~kpc. The Large and Small Magellanic Clouds are below the disk, with $z<0$. At the Sun's location, the positive $y$ axis is oriented counter to the direction of rotation of the disk.\footnote{This is equivalent to a $180^\circ$ rotation of the traditional Galactocentric coordinates used in \textsc{Astropy} \cite{astropy:2013,astropy:2018}.}

\subsection{Selecting a Complete RC/RGB Dataset}
We use the same dataset of high-luminosity stars within 4~kpc of the Solar location that we considered previously in Ref.~\cite{lim2023mapping}; this consists primarily of Red Clump (RC) and Red Giant Branch (RGB) stars. The RC is of particular interest in this work, as well as other recent studies (Refs.~\cite{2020A&A...643A..75S, 2024arXiv240608158B}) of equilibrium Galactic dynamics, since it is expected to be composed of relatively old stars. These old stars have had sufficient time to reach equilibrium, in the absence of any recent external dynamic perturbations such as mergers \cite{2014JPhG...41f3101R}.

We require that the stars in our sample have photometric magnitudes and small relative parallax errors: $\delta \varpi/\varpi < 1/3$. Although parallax errors in \Gaia{} DR3 are known to worsen beyond distances of several kpc, we do not observe any substantial spatial selection bias introduced by this cut.

To ensure that the sample is complete -- in the sense that a star within the sample would have been included in the dataset regardless of its distance from the Earth -- we require that every star in our sample is bright enough so that it would be above the \Gaia{} spectrometer magnitude threshold ${\rm G}_{\rm rvs} < 14$ if placed at the edge of the 4~kpc sphere of interest. The absolute magnitude (${M}_{G,{\rm rvs}}$, which we will denote as ${M}_{G}$) is calculated from the apparent magnitude ${\rm G}_{\rm rvs}$ and the distance modulus $\mu$:
\begin{equation}\label{eq:MG}
M_G = {\rm G}_{\rm rvs} - \mu(\vec{x}),
\end{equation}
where $\mu(\vec{x}) \equiv 5\log_{10}( |\vec{x}-\vec{x}_\odot|/{\rm kpc} )+10$, depending only on the distance between the star and the Sun's location $\vec{x}_\odot$. Note that this $M_G$ does not account for extinction. Our completeness criterion is then
\begin{equation} \label{eq:MGcompleteness}
M_G + \mu(4~{\rm kpc}) = M_G + 13.010 < 14.
\end{equation}

This criterion imposes an explicit cut on absolute magnitude $M_G<0.990$. The location of this cut with respect to prominent stellar populations in the \Gaia{} color-magnitude space is visualized in Figure~\ref{fig:cmd}. Prior to the selection on $M_G$, there are 24,789,061 well-measured stars in the \Gaia{} DR3 dataset within 4~kpc of the Sun which have full six-dimensional kinematics. This number is reduced to 5,811,956 by the magnitude selection, a majority of which ($\sim69\%$) lie within the RC.
The diagonal smearing of the color-magnitude diagram in Figure~\ref{fig:cmd} is caused by dust extinction. The vertical component of the smearing is due to the overall dimming of stars in the RVS band (increasing $M_G$), whereas the horizontal component of the smearing is due to ``reddening", the relative difference in dust extinction between the red and blue photometric bands of \Gaia{} (redder is increasing $BP-RP$).

We intentionally do not extinction-correct the primary dataset used in this work, as these corrections can only be applied to dimmed stars that are nearby enough for \Gaia{} to observe. Instead, we outline a novel technique for post-hoc dust correction in the next section.  In Section~\ref{sec:results}, we compare the machine-learned efficiency factor with the expectation from previous dust map studies. In particular, we use the dust map found in Ref.~\cite{Gaia2mass2022} to correct the $M_G$ of nearby stars to develop an independent expectation for the missing fraction of stars in nearby regions of space. We refer to this dust map as ``L22'' for the remainder of this work, and describe the construction of the efficiency function from L22 in Appendix~\ref{app:l22dust}.

Finally, we note that effects from crowding are clearly present in regions where the overwhelming flux of background starlight limits the ability of the \Gaia{} RVS to resolve individual stars. Most notably, the dark patch towards the Galactic center in Figure~\ref{fig:gaia_density_skymap} is due to Baade's window.

\section{Architecture and Training} \label{sec:training}

\subsection{MAF Models}\label{subsec:MAFs}

We train Masked Autoregressive Flows (MAFs, Ref.~\cite{papamakarios2018masked}) to learn the components of $f_{\rm obs}$ as well as their gradients:
\begin{equation}
    f_{\rm obs}(\vec{x},\vec{v})=n_{\rm obs}(\vec{x})p(\vec{v}|\vec{x})
\end{equation}
where $n(\vec{x})$ is the number density and $p(\vec{v}|\vec{x})$ is the conditional velocity distribution of the RC/RGB dataset. MAFs, a type of normalizing flow, model the density of data $p(x)$ by learning an invertible and differentiable transformation from a simple latent base distribution to $p(x)$. They are constructed from a composition of transformations that satisfy the autoregressive property, enabling a tractable Jacobian for the transformation.
This property allows us to efficiently compute gradients of $f_{\rm obs}$ with respect to $(\vec{x},\vec{v})$ by differentiating the MAFs directly with respect to their input parameters.
For a detailed discussion of normalizing flows, see Ref.~\cite{9089305,2019arXiv191202762P}.

\begin{figure}[t]
    \centering\includegraphics[width=0.6\columnwidth]{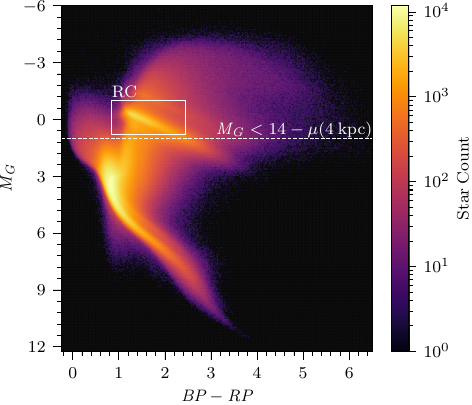}

    \caption{Absolute magnitude $M_G$ versus color $BP-RP$ of all \Gaia{} DR3 stars within 4~kpc of the Sun with full kinematic information, photometric magnitudes, and relative parallax errors $\delta \varpi/\varpi < 0.33$. The 4~kpc completeness cut is shown as a white dashed line: stars above this line outside the hatched region are used in our MAF training sample. The Red Clump (RC) -- the largest population used in this analysis -- is outlined in a white box. Figure reproduced from the authors' previous work, Ref.~\cite{lim2023mapping}.}
    \label{fig:cmd}
\end{figure}

The MAFs used in this work are similar, with minor updates, to those used in our previous work (Ref.~\cite{lim2023mapping}). The modifications are: 
\begin{enumerate}
\item 
Our data preprocessing has been updated to eliminate bias from the autoregressive structure of the MAF. As noted in Ref.~\cite{lim2023mapping}, MAFs are sensitive to the provided ordering of inputs (in our case, the ordering of Cartesian directions). To counteract this, we inserted Cartesian coordinate permutations between the layers of the MAF. However, since these permutations were identical across training realizations, it is possible that some residual autoregressive bias remained. In this work, we have introduced a further random rotation of the input $\{\vec{x},\vec{v}\}$ of the RC/RGB dataset after centering (${\overline{x}_i}'=0$) and standardization ($\sigma_{x_i}'=1$) for each training realization. 

\item In addition, the architecture of the MAFs has been updated to use $\rm{GINT}$
activations:
\begin{equation}
{\rm GINT}(x)=\frac{1}{\sqrt{2\pi}}\left(e^{-\frac{1}{2}x^2}-1\right)+\frac{1}{2}x\left(1+\rm{erf}\left(\frac{x}{\sqrt{2}}\right)\right).
\end{equation}
The GINT function is the integral of the Gaussian cumulative distribution function. As a result, its second derivative is the smooth Gaussian probability density function. This choice of activation function suppresses oscillations in the network, resulting in second-order derivative predictions which are more stable and less prone to fluctuations.
\end{enumerate}

Aside from these minor changes, the other details of the MAF models are unchanged, in particular they have the same size and structure as in Ref.~\cite{lim2023mapping}. The MAFs for both $n(\vec{x})$ and $p(\vec{v}|\vec{x})$ are composed of two MADE blocks, each with 10 hidden layers that are 48 nodes wide. All MAF models are trained by minimizing the negative log-likelihood (NLL) of the \Gaia{} DR3 RC/RGB dataset introduced in Section~\ref{sec:dataset}. For a complete discussion of the MAF training procedure, refer to Ref.~\cite{lim2023mapping}.

\subsection{Training Dust Efficiency and Potential}

We parameterize the functions $\Phi(\vec{x})$ and $\ln\epsilon(\vec{x})$ with fully-connected NNs. 
Both use five hidden layers, each 100 neurons wide, with the same ${\rm GINT}$ activations used for the MAFs. An additional transformation layer is included in the $\ln\epsilon$ network to smoothly map the output from $\mathbb{R}$ to $(-\infty,0)$ such that $\epsilon$ is constrained within $(0,1)$.\footnote{We choose the $C^\infty$ function $y=-\frac{1}{\pi}\left(x\left(\arctan{x}+\pi/2\right)+1\right)$ for this purpose.} The two networks are trained by minimizing the mean-square error (MSE) of CBE in Eq.~\eqref{eq:CBE_eps} with respect to the NN parameters $\bm{\theta}$ and $\bm{\vartheta}$:
\begin{equation}\label{eq:Lmseloss_reg}
\begin{aligned}
\mathcal{L}_{\bm{\theta} , \bm{\vartheta}}= \frac{1}{N_x N_v}\sum_{\vec{v}_{i,j} \sim p(\vec{v}|\vec{x_i})}^{N_{v}} \sum_{\vec{x}_i \sim n_{\rm obs}(\vec{x})}^{N_{x}} & \Biggl(\bigg| \vec{v}\,\cdot\vec{\nabla}\ln f_{\rm obs}(\vec{x},\vec{v}) - \vec{v}\,\cdot\vec{\nabla}\ln \epsilon_{\bm{\theta}}(\vec{x}) \\ 
& \left. \left.  - \vec{\nabla}\Phi_{\bm{\vartheta}}(\vec{x})\cdot \frac{\partial \ln f_{\rm obs}}{\partial \vec{v}\,} \right|^2 + \mathcal{L}_{\rm reg}(\vec{x})\right)_{\vec{x}_i,\vec{v}_{i,j}}
\end{aligned}
\end{equation}
where $\mathcal{L}_\textrm{reg}$ is a regularization term that imposes physical constraints on $\Phi$ and $\epsilon$. These regularization terms are discussed in more detail in the next subsection. Eq.~\eqref{eq:Lmseloss_reg} sums over $N_{x}=2^{22}\approx4\times10^{6}$ positions $\vec{x}_i$ sampled from the observed number density $n_\textrm{obs}(\vec{x})$, as well as $N_{v}=16$ velocities $v_{i,j}$ sampled at each $\vec{x}_i$ from $p(\vec{v}|\vec{x}_i)$, corresponding to $67.1$~million position-velocity pairs in total. 

We find that training using direct samples from the MAF serves as a form of importance sampling for well-constrained regions of phase space, yielding faster convergence than alternative sampling routines. MAF samples with spuriously high speeds ($|\vec{v}|>450\:\mathrm{km/s}$) are rejected, and sampling is repeated until $16$ valid velocity samples have been generated at every $\vec{x}$. Training is done in batches of $2^{14}$ samples, with 80\% of the total samples used for training and the remaining 20\% used to evaluate a validation loss.

Similar data preprocessing steps to those described in Section~\ref{subsec:MAFs} are used when training $\Phi$ and $\epsilon$ on these generated samples. Sampled data is centered, standardized, and then a random rotation is applied to both $\vec{x}$ and $\vec{v}$. During training, we solve the CBE in the regular, non-preprocessed space. Since $\Phi$ and $\epsilon$ take inputs in the preprocessed space, the appropriate Jacobian factors must be computed and included in Eq.~\eqref{eq:Lmseloss_reg} to correct their gradients.

We train in three stages: a ``coarse" stage with a learning rate of $10^{-2}$, and two ``refinement" stages with learning rates of $10^{-3}$ and $10^{-4}$ respectively. Each training stage ends when the validation loss does not improve after a predefined number of epochs called the ``patience". The patience for the first two stages is $25$, the final tuning stage has a shorter patience of $10$. After training completes, we select the epoch with the lowest validation loss.

\subsection{Regularization}\label{subsec:regularization}
Despite $\Phi(\vec{x})$ and $\epsilon(\vec{x})$ coupling to different gradients in the CBE, there is some remaining degeneracy between the solutions for the two NNs.
Therefore, we must encode some of our prior expectations for $\Phi$ and $\epsilon$ to push them towards a physical solution. To do so, we use the combination of two regularizers when minimizing Eq.~\eqref{eq:Lmseloss_reg}:
\begin{equation}
\mathcal{L}_{\rm reg}(\vec{x}) = \mathcal{L}_{\rm reg,\epsilon}(\vec{x})+\mathcal{L}_{\rm reg,\Phi}(\vec{x}),
\end{equation}
where $\mathcal{L}_{\rm reg,\epsilon}$ imposes an explicit scale for $\epsilon$ and $\mathcal{L}_{\rm reg,\Phi}$ penalizes $\Phi$ for regions of space with negative mass density. Although dust increasingly obscures stars along lines of sight from the Earth, our model of $\epsilon$ is not regularized or constrained to decrease monotonically with distance. This monotonicity is only guaranteed if all stars come from the same population of equal magnitude and color, whereas the RC/RGB dataset used in this work contains a combination of many stellar populations. In principle, stars behind a dust cloud could become intrinsically brighter or redder than the population of stars in front of the cloud, lessening the effect of the foreground dust. In practice, however, we find that $\epsilon$ naturally obeys this monotonicity in most (but not all) regions of the sky. Additionally, $\epsilon$ also corrects for other position-dependent sources of incompleteness such as crowding or mis-modeling by the MAF.

The form of Eq.~\eqref{eq:Lmseloss_reg} allows us to define the scale of $\epsilon$. Dust and crowding effects are mostly limited to the disk, and so we expect that $\epsilon \approx 1$ 
over most of the sky.
We encode this into our solution for $\epsilon$ by adding the following term to the MSE loss of Eq.~\eqref{eq:Lmseloss_reg}:
\begin{equation}
    \mathcal{L}_{\rm reg,\epsilon}(\vec{x}) = \lambda_\epsilon \left|\ln{\epsilon}(\vec{x})\right|^2.
    \label{eq:regLnEps}
\end{equation}
This regularization term penalizes solutions where $\epsilon \neq 1$ over large portions of the sky and diverges as $\epsilon$ approaches $0$, when $f_{\rm corr}(\vec{x})$ is no longer defined.
This term introduces an additional hyperparameter $\lambda_\epsilon$.
A hyperparameter scan over $\lambda_\epsilon$ reveals that small values of $\lambda_\epsilon$ fail to set a scale for $\epsilon$, leading to potentially inconsistent scales between different trainings of $\epsilon$ which will later be averaged together. Very large values of $\lambda_\epsilon$ force $\epsilon=1$ across the entire sky, a solution that fails to correct any of the effects of dust. Through numerical experiments, we found that the range $\lambda_\epsilon\in[10^{-2},1]$ yields stable results, and we ultimately settled on $\lambda_\epsilon=10^{-1}$ as the canonical value for this analysis.

Next, we introduce another regularization term penalizing negative mass density $\rho \propto \nabla^2 \Phi$ through the term
\begin{equation}
    \mathcal{L}_{\rm reg,\Phi}(\vec{x}) = \lambda_\Phi \textrm{ReLU}\left[-\nabla^2 \Phi(\vec{x})\right].
    \label{eq:regLnPhi}
\end{equation}
Ideally, this constraint would be satisfied exactly by construction in $\Phi$ instead of being regularized. As is discussed in the companion paper, some regions of space exhibit negative or near-zero mass densities. However, it is unclear that any equilibrium solution could be found in these regions if this is due to disequilibrium structures in the data. 

An additional benefit of regularizing $\Phi$ through its Laplacian is that it smooths $\Phi$, since noisy high-frequency features are explicitly penalized. However, introducing these noisy second derivatives in a loss with small batches of data can introduce significant training instability. Therefore, $\lambda_\Phi$ cannot be arbitrarily large. Numerical experiments showed that $\lambda_\Phi>\mathcal{O}(10)$ led to poor training stability, therefore we chose $\lambda_\Phi=10$ as the canonical value for this analysis.

\section{Results} \label{sec:results}
In this section, we present the results of the trained MAF model of $f_{\rm obs}$ as well as the $\epsilon$ network. In a companion paper Ref.~\cite{dustpaperII}, we analyze the associated dust-corrected $\Phi$ model, where we examine the resulting acceleration and mass density field of the local Milky Way.

\subsection{Dust Efficiency}

In order to assess whether $\epsilon$ is reasonably modeling incompleteness due to dust extinction, we directly compare it to the equivalent dust efficiency $\epsilon_{\rm L22}$ extracted from the extinction coefficients of the L22 dust map (see Appendix~\ref{app:l22dust} for details). Comparison between $\epsilon$ and $\epsilon_{\rm L22}$ is limited by several key factors:

\begin{enumerate}
    \item $\epsilon_{\rm L22}$ can only be computed where the RC/RGB subset of \Gaia{} is expected to be complete, regardless of dust extinction.
    This limits $\epsilon_{\rm L22}$ to distances of less than $d\sim2.5$~kpc, beyond which dust extinction drives stars below the \Gaia{} RVS detection threshold and cannot be recovered through extinction correction. \label{list:dust1}
    \item $\epsilon_{\rm L22}$ is computed from a \Gaia{} dust map in the monochromatic band at $5500$~\r{A}, whereas the RC/RGB sample of stars with 6D kinematic information were observed in the narrow $\sim8600$~\r{A} band of the \Gaia{} RVS. Depending on their chemical composition, dust clouds will have slightly different morphology when observed in different bands.
    \item $\epsilon$ corrects for all sources of disequilibrium in $f_{\rm obs}$, including true disequilibrium structures in the data or (to a limited extent) mismodeling of the data by the MAFs.
\end{enumerate}

\begin{figure*}[t!]
    \centering
    \begin{tabular}[t]{cc}
        \begin{tabular}[t]{c}
             \includegraphics[width=0.45\columnwidth]{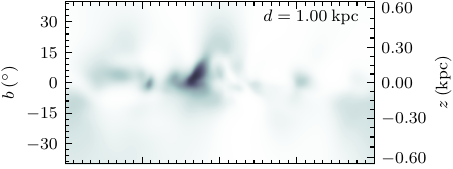}\\
             \includegraphics[width=0.45\columnwidth]{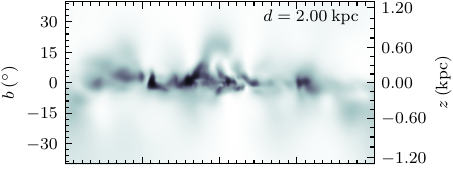}\\
             \includegraphics[width=0.45\columnwidth]{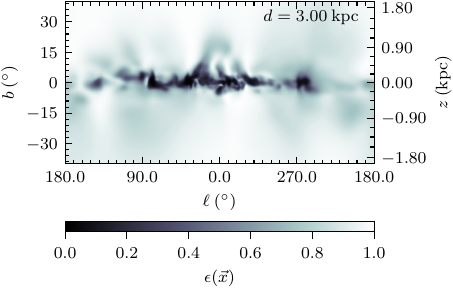}
        \end{tabular}&
        \begin{tabular}[t]{c}
             \includegraphics[width=0.45\columnwidth]{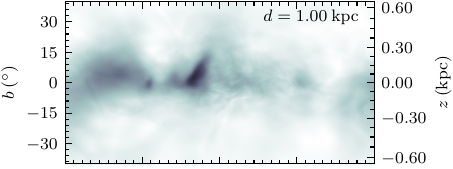}\\
             \includegraphics[width=0.45\columnwidth]{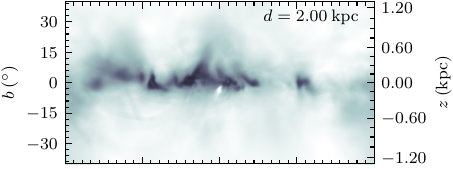}\\
             \includegraphics[width=0.45\columnwidth]{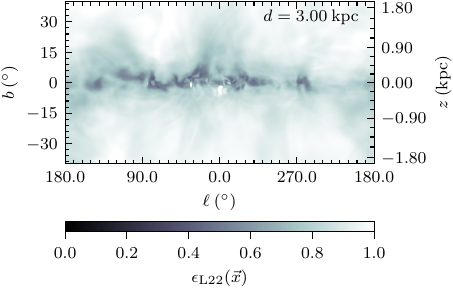}
        \end{tabular}
    \end{tabular}
    \caption{Left: CBE-derived $\epsilon$ projected in Galactic coordinates at distances $d=1$, $2$, and $3$~kpc. At larger distances, familiar features in local dust clouds become more resolved in regions where $\epsilon<1$. However, not every feature seen here should be attributed to dust extinction. For example, the dark patch that first appears at $(\ell,b,d)\approx(5^{\circ},-4^{\circ},2\:\textrm{kpc})$ is an incomplete patch in the \Gaia{} dataset due to crowding and saturation effects. Right: Same as left, using $\epsilon_{\rm L22}$ derived from the L22 dust map \cite{Gaia2mass2022}. At $3$~kpc, $\epsilon_{\rm L22}$ approaches $1$ and is no longer reliable, as it cannot restore stars dimmed beyond the limits of the \Gaia{} spectrometer.
    \label{fig:compared_epsilon_skymaps}}
\end{figure*}

We proceed with the comparison with these limitations in mind. In Figure~\ref{fig:compared_epsilon_skymaps}, we benchmark the learned $\epsilon$ against $\epsilon_{\rm L22}$ across the sky at different distances from the Earth.\footnote{Although we include the $d=3$~kpc panel for $\epsilon_{\rm L22}$ in Fig.~\ref{fig:compared_epsilon_skymaps}, it is noticeably degraded in quality compared to the other panels at $d=1$ and $d=2$~kpc. This is to be expected from point~\ref{list:dust1} above: the distance limitation on the $\epsilon_{\rm L22}$ model means we cannot reliably benchmark $\epsilon$ beyond $d=2.5$~kpc.} The $\epsilon$ model detects all of the major dust clouds highlighted in Figure~\ref{fig:gaia_density_skymap}, and appears to accurately model the dense complex of dust towards the Galactic center. However, there is minor disagreement between $\epsilon$ and $\epsilon_{\rm L22}$. As is evident in the top panel of Figure~\ref{fig:compared_epsilon_skymaps} at $1$~kpc, our model for $\epsilon$ underestimates the amount of dust extinction at small distances. This has the effect of ``overestimating" the distance to some nearby dust clouds, most prominently in the direction of Cygnus and CEP OB3. Some disagreement between $\epsilon$ and $\epsilon_{\rm L22}$ is expected. For example, the small dark patch visible in $\epsilon$ near $\ell\approx0^\circ$ and $b\approx-3^\circ$ appears as a bright white patch in $\epsilon_{\rm L22}$. Here, $\epsilon$ is learning to correct for the crowding near Baade's window, which is not an extinction effect from dust (hence $\epsilon_{\rm L22}\approx1$).

\begin{figure*}[t]
    \centering
    \includegraphics[width=0.93\columnwidth]{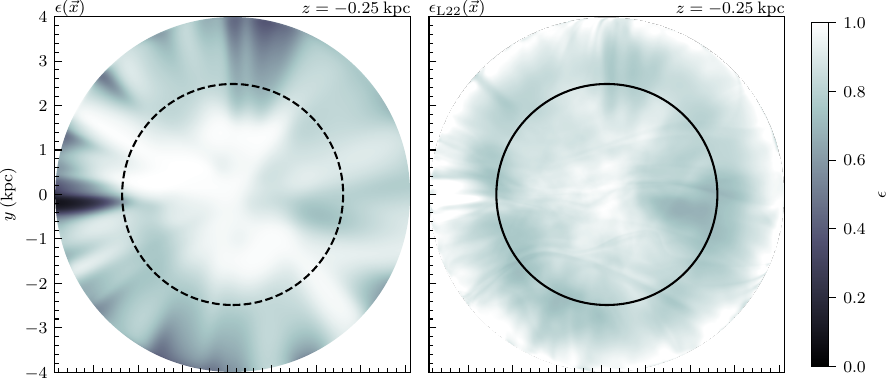}
    \includegraphics[width=0.93\columnwidth]{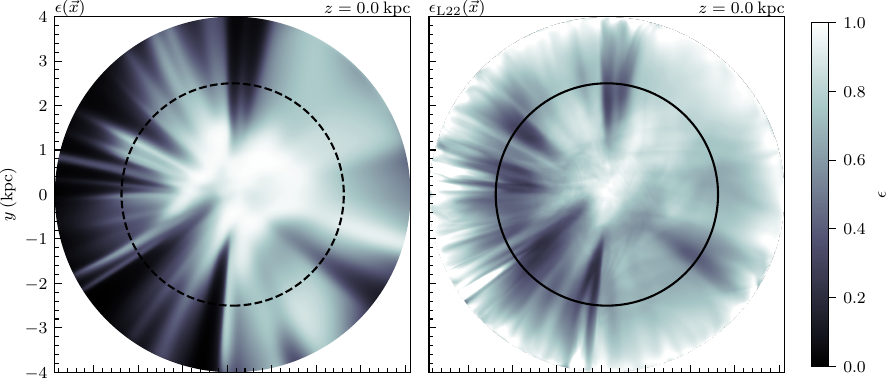}
    \includegraphics[width=0.93\columnwidth]{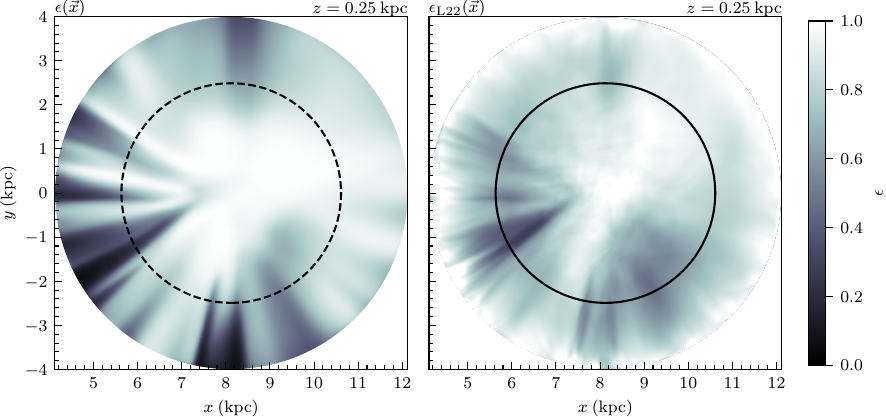}

    \caption{Left: Dust efficiency function $\epsilon$ estimated from the CBE along slices in the $x-y$ plane at $z=-0.25$ (top), $0.0$ (center), and $0.25$~kpc (bottom). The center row is a slice through the midplane of the Milky Way's disk, where the most efficiency loss is expected due to the dense network of dust clouds.
    Right: $\epsilon_\textrm{L22}$ estimated from the extinctions provided by the L22 dust map, limited to within $d=2.5$~kpc (black circle). Beyond $d=2.5$~kpc, $\epsilon_{\rm L22}$ cannot restore missing stars and returns to $1$.}
    \label{fig:epsilon_skymaps_plane}
\end{figure*}

To understand the structure of $\epsilon$ both deep within the Milky Way's disk as well as on the disk's outskirts, we examine $\epsilon$ within tomographic slices through space at fixed $z$ in Figure~\ref{fig:epsilon_skymaps_plane}. In the outer disk at $|z|=0.25$~kpc, minor dust lanes visible in $\epsilon_{\rm L22}$ are recovered in $\epsilon$. At larger radii, beyond the reach of our ${\rm L22}$ estimate of $\epsilon$, significant dust lanes near the Galactic center begin to appear. Within the center of the disk, shown in the central panel of Figure~\ref{fig:epsilon_skymaps_plane}, we observe a complex structure of nearby dust clouds, which all mostly begin to appear within $0.5-2$~kpc of the Earth. Interestingly, while most rays show monotonically decreasing $\epsilon$ with distance, select regions show $\epsilon$ slightly increasing with distance.

\begin{figure*}
    \centering
    \includegraphics[width=0.9\columnwidth]{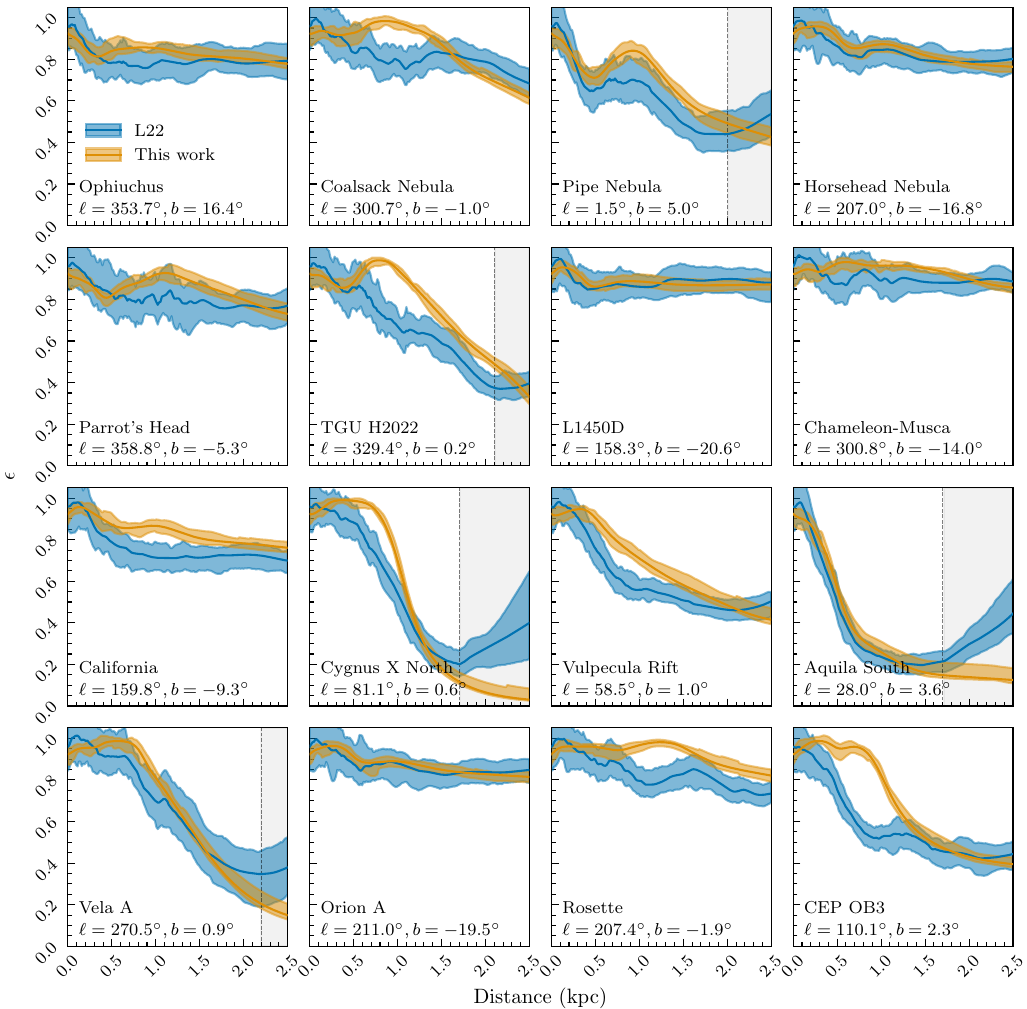}
    \caption{Comparison of $\epsilon_{\rm L22}$ derived from Ref.~\cite{Gaia2mass2022} (blue) and the $\epsilon$ estimated in this work (gold) along lines of sight centered on known significant dust clouds shown in Figure~\ref{fig:gaia_density_skymap}. The center line represents an average over a cone of opening angle $\theta=10$~arcmin. The error band for the $\epsilon$ model (blue) is dominated by measurement and statistical uncertainties. Likewise, the error band of the $\epsilon_{\rm L22}$ model (gold) includes the L22 dust map errors as well as an estimate of measurement and statistical errors. Both error bands include the subdominant variation across each circular section of the cone. The light gray band in some panels indicates lines of sight where $\epsilon_{\rm L22}$ degrades closer than $d=2.5$~kpc due to heavy dust extinction.}
    \label{fig:epsilon_comparison_cloud}
\end{figure*}

In Figure~\ref{fig:epsilon_comparison_cloud}, we next examine $\epsilon$ evaluated within cones of radius $10$~arcmin that pass through the sixteen dusty regions of the Solar neighborhood highlighted in Figure~\ref{fig:gaia_density_skymap}.
The uncertainties in $\epsilon$ shown in Figure~\ref{fig:epsilon_comparison_cloud} are derived by repeatedly retraining the MAFs and solving the CBE using multiple measurement error-smeared or resampled copies of the \Gaia{} data, and by computing the variation across hundreds of retrainings with different network initializations. A full description of the uncertainty quantification procedure is given in Appendix~\ref{app:uncertainties}. The error band for $\epsilon_{\rm L22}$ is similarly derived, using the provided uncertainties in the L22 dust map \cite{Gaia2mass2022} as well as error-smeared and resampled copies of the \Gaia{} data.
Additionally, for both $\epsilon$ and $\epsilon_{\rm L22}$, we take the average and variance over a short angular patch or ``cone" in order to mitigate differences in resolution between the models. Variance across the cone is found to be subdominant to the primary measurement and statistical uncertainties.

In all but two cases in Figure~\ref{fig:epsilon_comparison_cloud}, the $\epsilon$ and $\epsilon_{\rm L22}$ are the same within errors at large distances $(1.5{\rm\:kpc}<d<2.5{\rm\:kpc})$. At intermediate distances $(0.5{\rm\:kpc}<d<1.5{\rm\:kpc})$, our model of $\epsilon$ underestimates the amount of extinction relative to $\epsilon_{\rm L22}$ beyond $1\sigma$ uncertainties in seven cases: Coalsack, TGU H2022, California, Cygnus, Vulpecula, Rosette, and CEP OB3. Pronounced bumps are visible in $\epsilon$ along several lines of sight at $d\sim 0.9$~kpc. It is possible that this is an artifact of \Gaia{} parallaxes, which deteriorate in quality with distance. Alternative distance measures are typically used at larger distances, however these are not uniformly available across the sky and are therefore unsuitable for this analysis. In CEP OB3 and Cygnus, the start of the turndown in $\epsilon$ is delayed relative to $\epsilon_{\rm L22}$. It is unclear if this is due to some incompatibility between the monochromatic and RVS bands, or some other form of mismodeling, but it follows a general trend where $\epsilon$ appears to overestimate the distance to the beginning of dust clouds relative to $\epsilon_{\rm L22}$.

\begin{figure}
    \centering
    \includegraphics[width=\columnwidth]{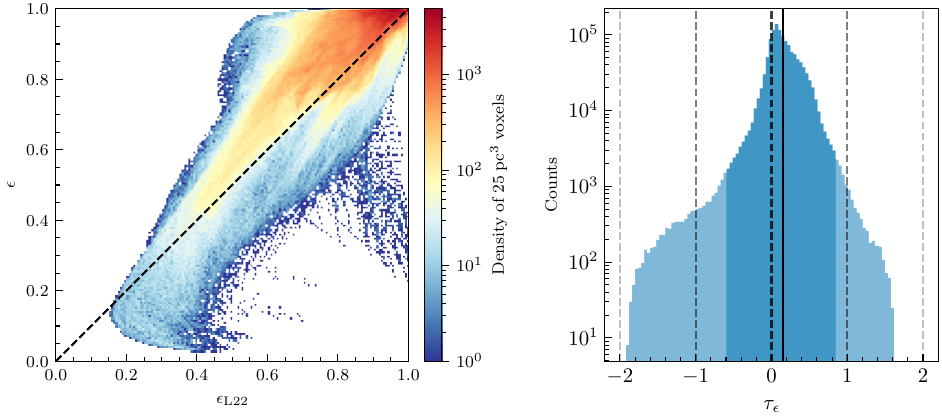}
    \caption{Pointwise ``voxel" comparison between $\epsilon_{\rm L22}$ derived from Ref.~\cite{Gaia2mass2022} and the $\epsilon$ model that solves Eq.~\eqref{eq:CBE_eps}. Each efficiency value is evaluated over a grid of 3D voxels of volume $(25\:{\rm pc})^3$ that tile the space within $|z| <0.5$~kpc of and $d<2.5$~kpc. Top: Frequency of efficiency values in each voxel. 
    Color scale indicates the density of voxels in each bin in a logarithmic scale. Bottom: Frequency of $\tau_\epsilon$ values between $\epsilon$ and $\epsilon_{\rm L22}$ accounting for errors. Dark blue indicates the $1-99\%$ percentile range of pull values. The solid black line indicates the average pull value.}
    \label{fig:epsilon_comparison_2d_plot}
\end{figure}

As a final test, we compare $\epsilon$ and $\epsilon_{\rm L22}$ pointwise across a uniform, three-dimensional grid of nearby voxels in the Milky Way's disk ($d<2.5$~kpc and $|z|<0.5$~kpc). In the left panel of Figure~\ref{fig:epsilon_comparison_2d_plot}, we find that the models agree when there is no predicted extinction ($\epsilon\approx 1$). Additionally, we observe that $\epsilon$ lags behind as $\epsilon_{\rm L22}$ decreases from 1. This lag supports our previous observations that $\epsilon$ decreases from $1$ slightly later than $\epsilon_{\rm L22}$ along a given line of sight. This disagreement is slight, on the order of $||\epsilon-\epsilon_{L22}||\approx0.1$ when both $\epsilon$ and $\epsilon_{\rm L22}>0.4$. In practice, at a given $\vec{x}$ within a region of moderate dust extinction, this means that these models on average disagree at the 10\% level on the fraction of missing stars. This is consistent with the level of disagreement seen in Figure~\ref{fig:epsilon_comparison_cloud}. In the right panel of Figure~\ref{fig:epsilon_comparison_2d_plot}, we present the pull
$\tau_\epsilon=(\epsilon-\epsilon_{\rm L22})/\sqrt{\sigma_\epsilon^2+\sigma_{\epsilon_{\rm L22}}^2}$
between $\epsilon$ and $\epsilon_{\rm L22}$ evaluated at each voxel when accounting for uncertainties. As can be seen from the pull distribution, the bulk of $\tau_\epsilon$ values skew slightly positive, indicating that the amount of dust extinction tends to be slightly underestimated relative to the L22 dust map. However, the absolute value of this pull rarely exceeds $1\sigma$, indicating excellent agreement between these two models.

\subsection{Dust-Corrected PSD}

\begin{figure*}
    \centering
    \begin{tabular}[t]{cc}
        \begin{tabular}[t]{c}
        \includegraphics[width=0.45\columnwidth]{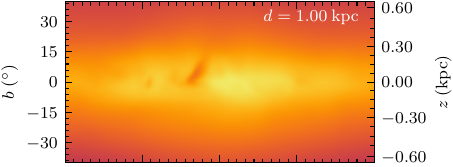}\\
        \includegraphics[width=0.45\columnwidth]{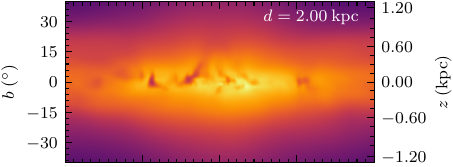}\\
        \includegraphics[width=0.45\columnwidth]{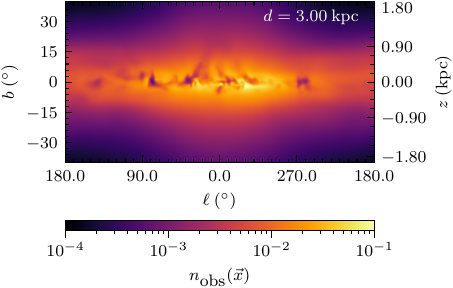}
        \end{tabular}&
        \begin{tabular}[t]{c}
        \includegraphics[width=0.45\columnwidth]{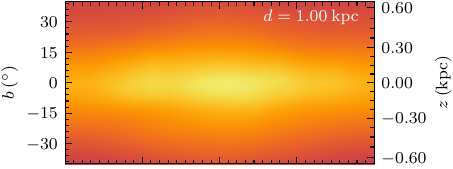}\\
        \includegraphics[width=0.45\columnwidth]{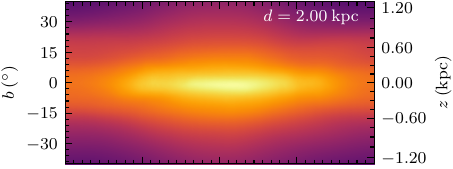}\\
        \includegraphics[width=0.45\columnwidth]{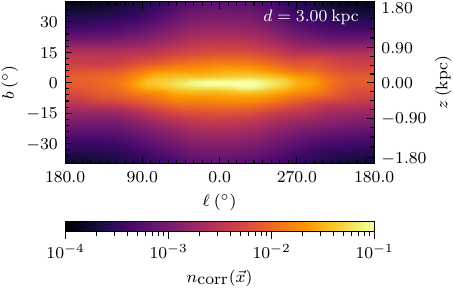}
        \end{tabular}
    \end{tabular}
    \caption{Left: MAF model of the observed number density $n_\textrm{obs}$ of \Gaia{} RC/RGB stars projected in Galactic coordinates at distances $d=1$ (top), $2$ (center), and $3$~kpc (bottom). Right: Extinction-corrected number density $n_{\rm corr} \equiv n_\textrm{obs}/\epsilon$.}
   \label{fig:corrected_gaia_flow_density}
\end{figure*}

\begin{figure*}
\centering
\includegraphics[width=\columnwidth]{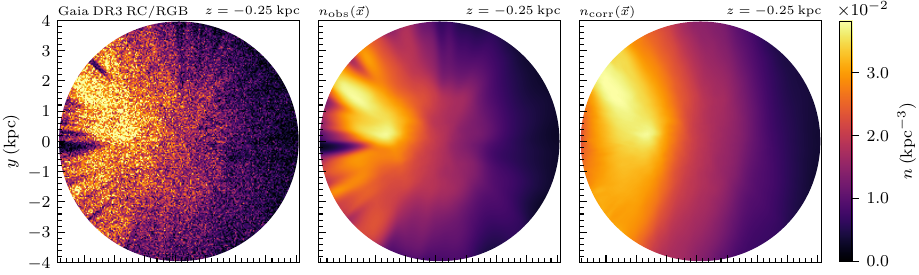}
\includegraphics[width=\columnwidth]{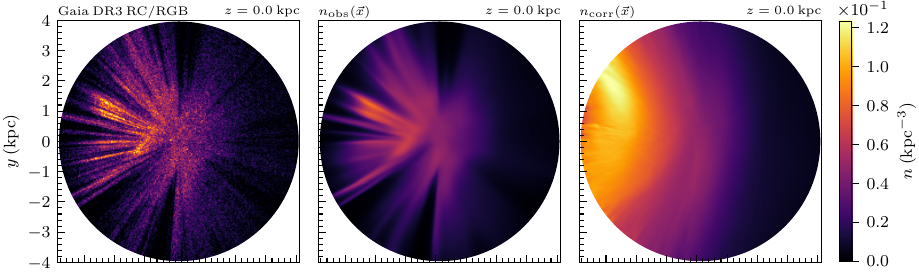}
\includegraphics[width=\columnwidth]{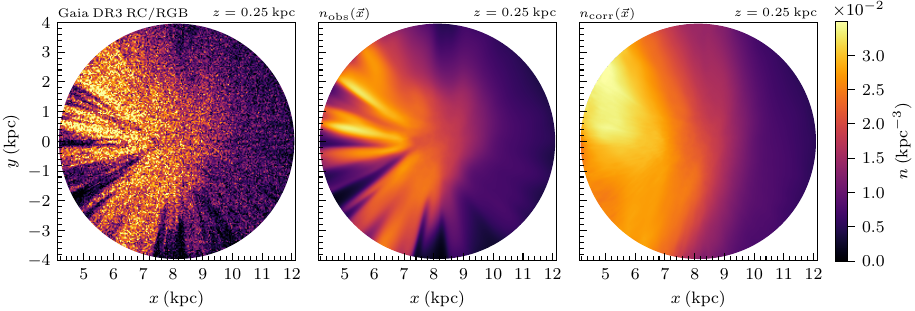}
\caption{Left: Histogram of the \Gaia{} RC/RGB data (see Section~\ref{sec:dataset}) binned in $37.5$~pc cubes along slices in the $x-y$ plane at $z=-0.25$ (top), $0$ (center), and $0.25$ (bottom) kpc. Center: MAF model of the observed number density $n_{\rm obs}$ of stars evaluated over the same slices in the $x-y$ plane at intervals of $12.5$~pc. Right: Dust-corrected model of the number density $n_{\rm corr}$ evaluated at the same points as $n_{\rm obs}$. In each row, the color scale is set to the $99^{\rm th}$ percentile value of the plotted $n_{\rm corr}$.}
\label{fig:corrected_tomographic_slices}
\end{figure*}

Using Eq.~\eqref{eq:epsilon_def}, we apply the learned dust efficiency $\epsilon$ to dust-correct the MAF-estimated $f_{\rm obs}$, yielding the dust-corrected PSD $f_{\rm corr}$ up to an overall normalization constant. In the right panel of Figure~\ref{fig:corrected_gaia_flow_density}, we present the corrected stellar number density $n_{\rm corr}$ in slices of constant distance across the sky. Compared to $n_{\rm obs}$ in the left panel of Figure~\ref{fig:corrected_gaia_flow_density}, phase space suppression from dust-extinction is almost completely removed by our $\epsilon$ correction procedure up to distances of $3$~kpc from the Solar location.

This novel dust correction technique provides us with a much-improved PSD and solution to the CBE. In Eq.~\eqref{eq:CBE_eps}, the potential $\Phi$ enters at the same order as $f_{\rm corr}$, and so we expect an accurate $f_{\rm corr}$ to yield an accurate $\Phi$. However, downstream quantities such as the gravitational acceleration $\vec{a}$ and mass density $\rho$ enter at the same order as the first- and second-order gradients of $f_{\rm corr}$ respectively, and are therefore much more sensitive to mismodeling of $f_{\rm corr}$. Although analysis of $\vec{a}$ and $\rho$ are reserved for our companion paper Ref.~\cite{dustpaperII}, we now assess our dust-correction procedure in the most challenging region of space: in the disk towards the Galactic center.

The primary obstacle towards accurate modeling of $\epsilon$ in this volume of our data comes from the large number and close proximity of dust clouds. These limit the number of nearby clear lines of sight that the $\epsilon$ model can use to ``interpolate" into dusty regions. This is exacerbated towards the boundary of the $4$~kpc window ($x \lesssim 4.5$~kpc), as $\epsilon$ has limited information from the stellar halo.

To examine the quality of our corrected phase space density in this challenging region of space, we show $n_{\rm corr}$ alongside $n_{\rm obs}$ in Figure~\ref{fig:corrected_tomographic_slices} along tomographic slices in the $x-y$ plane in the disk. At $z=0$, exactly in the midplane, the expected structure of the disk emerges in $n_{\rm corr}$ from the heavily obscured $n_{\rm obs}$. Note that this is a purely data-driven dynamical inference of the structure of the Milky Way's disk, as the $\epsilon$ network was not given any inductive bias regarding the structure of the Galactic disk. However, we can see in Figure~\ref{fig:corrected_tomographic_slices} likely-spurious residual structures, both at small scales and large. Examples of the small-scale mismodeling include the ``fringing'' seen in $n_{\rm corr}$ at $z=0$ and $x\sim 5$~kpc, and an unexpected decrease in $n_{\rm corr}$ very close to the edge of the observation volume towards the Galactic center. At larger scales, there is a notable asymmetry in the overall number density of stars between $y>0$ and $y<0$ at small $x$. The fringing in particular is likely to affect the quality of the downstream inference for $\vec{a}$ and $\rho$, discussed in \cite{dustpaperII}.

There are several possible sources of these errors. Fundamentally, the data towards the Galactic center within the disk is characterized by narrow lines of unclouded sight towards the Galactic center (seen in the left column of Figure~\ref{fig:corrected_tomographic_slices}). As a result, much of this region has nearly no stars in $n_{\rm obs}$. Regressing $\epsilon$ -- the ratio between $n_{\rm corr}$ and $n_{\rm obs}$ becomes challenging when the denominator is nearly zero. The $\epsilon$ network may also lack sufficient expressivity to fully correct small-scale features in the observed phase space -- note that the fringing appears to correlate with the edges of the clear lines of sight.

\begin{figure}[t]
    \centering
    \includegraphics[width=\columnwidth]{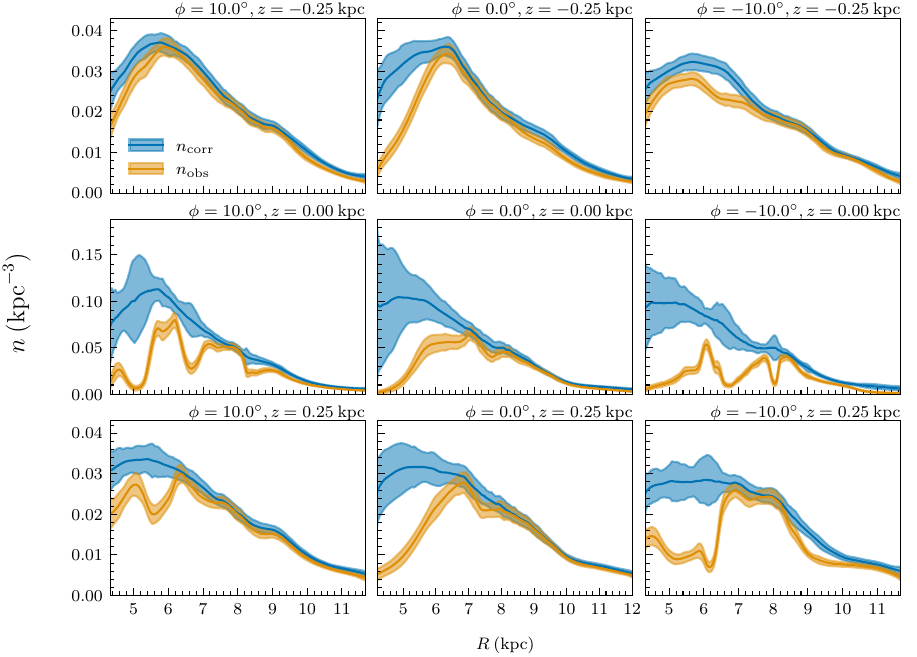}
    \caption{$n_{\rm corr}$ (blue) and $n_{\rm obs}$ (gold) with their $1\sigma$ uncertainties evaluated along scan lines in Galactic cylindrical radius $R$ originating from $(x,y) = (0,0)$ in the directions $\phi=(10,0,-10)^\circ$ (left, center, and right), each at $z=(-0.25, 0, 0.25)$~kpc (top, center, bottom).}
    \label{fig:ncorr_errs}
\end{figure}

We note that the two-dimensional plots in Figure~\ref{fig:corrected_tomographic_slices} reflect the mean output of the observed and corrected MAFs. This obscures the uncertainties inherent in our results. In Figure~\ref{fig:ncorr_errs}, we plot the $n_{\rm obs}$ and $n_{\rm corr}$ as a function of cylindrical radius $R$ along rays emanating from $(x,y) = (0,0)$ (at $\phi = -10^\circ$, $0^\circ$, $+10^\circ$, and $z = -0.25$, $0$, and $+0.25$~kpc) including the full error budget as described in Appendix~\ref{app:uncertainties}.  
As can be seen, the errors in $n_{\rm corr}$ increase towards the Galactic center and within the disk, indicating the lack of confidence the $\epsilon$ network has in a particular solution as the data becomes sparser and regression of $\epsilon$ more difficult. This is the expected and desired behavior: similar uncertainties will propagate into results for $\Phi$ and downstream observables, avoiding overconfident predictions unsupported by the data. However, as an important exception to this, we do note that in some regions (particularly at $z=-0.25$~kpc) the decrease in $n_{\rm corr}$ towards the Galactic center remains statistically significant even as the errors increase. This behavior is unexpected, and may reflect modeling issues at the boundary of our data volume that have not been fully accounted for in the error budget.

Apart from these issues towards the Galactic center, the plots in Fig.~\ref{fig:ncorr_errs} further illustrate the impressive successes of our technique in smoothing out the many wiggles in $n_{\rm obs}$ caused by dust extinction, and generally recovering azimuthal and north-south symmetry that is to be expected from an equilibrium stellar population. We emphasize again that these symmetries were not put into our analysis by hand, but emerged in a fully data-driven way from our analysis.

\section{Discussion} \label{sec:discussion}

We have developed a novel dust correction technique for astrometric datasets of equilibrated or approximately-equilibrated stellar tracer populations. By simultaneously training neural networks that model the dust efficiency and Galactic potential using the CBE as a loss, we can estimate the underlying true PSD of stars in regions of the Galactic disk that are almost completely obscured by dust. In this paper, we described the algorithm and the training, along with investigating the resulting dust efficiency map. In our companion work \cite{dustpaperII}, we show results for the $\Phi$ network, extracting a three-dimensional map of the gravitational potential, accelerations, and mass density across our $4$~kpc analysis volume.

Overall, the dust efficiency field $\epsilon(\vec{x})$ presented in this work is in excellent agreement with a recent high-resolution 3D dust map (Ref.~\cite{Gaia2mass2022}). However, some discrepancies and challenges remain, especially in the disk towards the Galactic center where there are many regions of heavy dust extinction with extremely sparse data.
The $\epsilon$ network observed an increasing density along some of the few clear lines of sight towards the Galactic center, but it did not infer that the density should increase along other nearby lines of sight. This led to a decrease in $n_{\rm corr}$ below Galactic radii of $R<5$~kpc, as well as a non-axisymmetric structure in $n_{\rm corr}$ within $5<R<6$~kpc. However, we note that most of these unexpected features are not significant when uncertainties are considered.

While $n_{\rm corr}$ could be constrained to monotonically increase towards the Galactic center, the most model-independent way to resolve these issues is to use a larger dataset of stars and to move the boundary of the observation window further away. We anticipate greater availability of 6D data in future \Gaia{} data releases, and it may also be possible to use information from other surveys to obtain an improved data-driven estimate of the true phase space density at the Galactic center.

These challenges could also potentially be mitigated with future improvements to our technique. These include employing more advanced density estimation methods, such as diffusion models \cite{song2021scorebased} or flow matching \cite{lipman2023flow, tong2024improving}, or learning $f$ using Fourier features \cite{10.5555/3495724.3496356} to directly address the small-scale spectral bias \cite{pmlr-v97-rahaman19a}. Another potential approach would involve modeling $f_{\rm corr}$ with a MAF and training $f_{\rm corr}$, $\epsilon$, and $\Phi$ in a single step by minimizing the CBE and the NLL simultaneously. Since $f_{\rm corr}$ is inherently smoother than $f_{\rm obs}$, this method would allow the MAF to focus less on modeling high-frequency dust-related features. While promising, these alternatives would require new hyperparameters and training regimens, which we leave for future work.

Finally, we emphasize that the $\epsilon$ presented in this work serves a unique role compared to traditional dust maps. Instead of providing the magnitude of dust extinction, $\epsilon$ provides a direct estimate of the fraction of stars that dust extinction (or other effects, such as crowding) has removed from the sample. This serves as a direct estimate of the \Gaia{} selection function in the RVS band. However, we caution that it is not possible to convert from $\epsilon$ to a traditional dust extinction map. Although we were able to convert the ${\rm L22}$ dust map into an effective $\epsilon_{\rm L22}$, this was only possible in limited regions of space where the \Gaia{} survey was expected to be complete. Instead, we present $\epsilon$ as a parallel technique for mapping the effects of local dust clouds, not as a replacement for the available highly detailed 3D dust maps.

\acknowledgments

This work was supported by the DOE under Award Number DOE-SC0010008. 
The work of SHL was also partly supported by IBS under the project code, IBS-R018-D1. 
This work was also performed in part at Aspen Center for Physics, which is supported by National Science Foundation grant PHY-2210452. 
The authors acknowledge the Office of Advanced Research Computing (OARC) at Rutgers, The State University of New Jersey for providing access to the Amarel cluster and associated research computing resources that have contributed to the results reported here. URL: \url{https://oarc.rutgers.edu}.
This research used resources of the National Energy Research Scientific Computing Center, a DOE Office of Science User Facility supported by the Office of Science of the U.S. Department of Energy under Contract No. DE-AC02-05CH11231 using NERSC award HEP-ERCAP0027491.

\appendix

\section{Velocity Independence of the Dust Efficiency} \label{app:eps_velo}

In this work, we have relied on the crucial assumption that the dust efficiency factor is velocity-independent, see Eq.~\eqref{eq:epsilon_def}. Here we will further justify this assumption and provide evidence for it from \Gaia{} DR3 data. 

As discussed above, the PSD can be decomposed into a number density and a conditional velocity distribution:
\begin{equation}
f(\vec x,\vec v)=n(\vec x)p(\vec v|\vec x)
\end{equation}
We are interested in the difference between the observed and dust-corrected PSDs; the ratio between the two is the dust efficiency factor:
\begin{equation}
\epsilon(\vec x,\vec v)=\frac{n_{\rm obs}(\vec x)p_{\rm obs}(\vec v|\vec x)}{n_{\rm corr}(\vec x)p_{\rm corr}(\vec v|\vec x)}
\end{equation}
On general grounds, we expect that the velocity distribution is unaffected by dust extinction along the line of sight {\it for stars of a given magnitude}, i.e., 
\begin{equation}
\label{eq:pobspcorr}
p_{\rm obs }(\vec{v}|\vec{x}, M_G) = p_{\rm corr}(\vec{v}|\vec{x}, M_G)
\end{equation}
If all of our stars were of the same magnitude, this would be enough to guarantee velocity independence of $\epsilon$. However, in general, our data is composed of a superposition of stars with different magnitudes
\begin{equation}\label{eq:epsilon_definition_1}
    \epsilon(\vec{x}, \vec{v}) = \frac{\int dM_G \, n_{\text{obs}}(\vec{x}|M_G) \, p_{\text{obs}}(\vec{v}|\vec{x}, M_G) \, p_{\text{obs}}(M_G)}{\int  dM_G \, n_{\text{corr}}(\vec{x}|M_G) \, p_{\text{corr}}(\vec{v}|\vec{x}, M_G) \, p_{\text{corr}}(M_G)}
\end{equation}
and dust extinction can affect these differently. For example, dimmer stars may be driven below the \Gaia{} observability threshold by dust extinction, while brighter stars may be left in the sample even after dust extinction.
This can lead to different observed vs.\ corrected number densities,
and a residual velocity dependence of $\epsilon$ in Eq.~\eqref{eq:epsilon_definition_1}. However, if the probability of a star at a given $\vec{x}$ to have velocity $\vec{v}$ is independent of $M_G$ (for both the observed and corrected datasets), then $\epsilon$ will depend only on $\vec{x}$.

Since the stars we consider in this work are primarily RC/RG stars that are located near the disk, they are predominately drawn from a single population -- older disk stars \cite{2021ApJ...916...88G}. Assuming this population is in equilibrium, then the velocity distribution at every location is expected to be the same for all stars of this population; and so we should expect $p(\vec{v}|\vec{x},M_G)$ to be independent of $M_G$ and thus that the efficiency factor is velocity-independent. 

\begin{figure}
    \centering
    \includegraphics[width=.95\columnwidth]{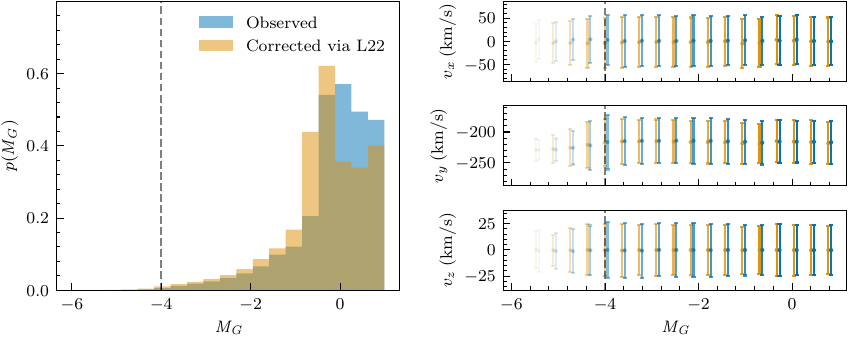}
    \caption{
    Left: Distribution of $M_{G}$ in the observed (blue) and dust-corrected (gold) datasets. Right: Mean (center point) and standard deviation (error bar) of each Cartesian component of the velocity distribution of the observed and dust-corrected datasets, binned by $M_G$. Opacity of points is set by $\log\left(p(M_G)\right)$.}
    \label{fig:pMG_hist_dispersion_scatter}
\end{figure}

Support for this expectation can be found through examination of the \Gaia{} DR3 data. We build a secondary dataset of stars within $2$~kpc that are nearby enough to be complete up to dust-corrected magnitudes of $M_G+\mu+A_{\rm rvs} \leq 14$, using $A_{\rm rvs}$ computed from the L22 dust map (Ref.~\cite{Gaia2mass2022}). This differs from our primary dataset, which uses the cut in Eq.~\eqref{eq:MGcompleteness} that instead relies on uncorrected magnitudes. The distribution of $M_G$ for stars within this dataset (both before and after correcting for dust-extinction) are shown in the left panel of Figure~\ref{fig:pMG_hist_dispersion_scatter}. In the right panel of this Figure, we show the mean and standard deviations of the distributions of each of the three components of the stellar velocities (in Galactocentric Cartesian coordinates), binned in $M_G$. Both before and after correcting for dust, it is apparent that for stars with $M_G \gtrsim -4$, the probability distribution of velocities is largely independent of $M_G$. This matches our expectation if the majority of stars are from a single population, and is the necessary condition for the efficiency factor to depend only on position.

\section{Estimating an Efficiency from the L22 Dust Map} \label{app:l22dust}

In this appendix, we introduce a technique for inverting an existing dust map (L22, Ref.~\cite{Gaia2mass2022}) to a reference selection function $\epsilon_{\rm L22}$ with which we can use to validate our CBE-derived selection function $\epsilon$. Note that $\epsilon_{\rm L22}$ is only used as a benchmark, and is not used in any of our downstream analyses of the estimated dust-corrected PSD or gravitational potential. L22 provides a 3-dimensional map of monochromatic extinctions $(A_0)$ at a reference wavelength of $5500$~\r{A} within our entire region of interest. The passband of the \Gaia{} RVS is relatively narrow, so monochromatic extinctions can be converted to extinction in the RVS band ($A_{\rm rvs}$) using a simple extinction law\footnote{https://www.cosmos.esa.int/web/gaia/edr3-extinction-law}:
\begin{equation}
    A_{\rm rvs}=k_{\rm rvs}A_0
\end{equation}
where $k_{\rm rvs}=0.5385$ \cite{2023A&A...674A...6S}. Note that this conversion cannot account for relative differences in extinction from structures with different absorption at $\lambda_0=5500$~\r{A} versus the RVS band at $\lambda_{\rm RVS}\approx8550$~\r{A}. This limits comparisons between our estimate of $\epsilon$ and $\epsilon_{\rm L22}$ inferred from the L22 dust map, as both trace structures in the local dust field observed in different bands. Consequently, there will be some level of inherent incompatibility between these two selection functions in some regions of space or at smaller length scales.

\begin{figure}[h]
    \centering \includegraphics[width=0.6\columnwidth]{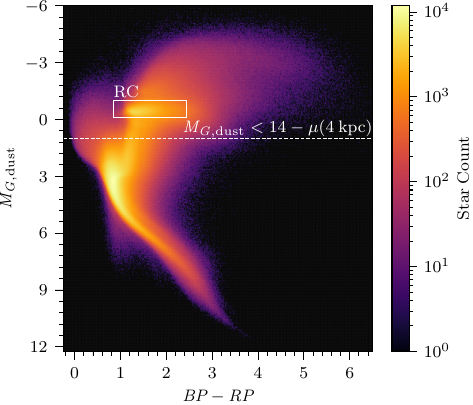}

    \caption{Same as Figure~\ref{fig:cmd}, using extinction-corrected absolute magnitudes $M_{G,{\rm dust}}$ computed from the L22 dust map (Ref.~\cite{Gaia2mass2022}).}
    \label{fig:cmd_L22}
\end{figure}

In Figure~\ref{fig:cmd_L22}, we show the colors and magnitudes of the stars in our 4~kpc data sphere, corrected by the L22 dust map. Compared to Figure~\ref{fig:cmd}, the RC/RBG is more concentrated in color and magnitude after the dust correction is applied, and the overall diagonal smearing visible across Figure~\ref{fig:cmd} is no longer present. This indicates that the L22 dust correction appropriately clusters individual populations in ${\rm G}_{\rm RVS}$-space and is properly correcting for dust extinction.

We train an additional set of MAF models to learn the number density $n_{\rm L22}$ of stars with dust-corrected magnitudes $M_{G,{\rm dust}}$ bright enough to satisfy Eq.~\eqref{eq:MGcompleteness}, i.e. the stars above the white dashed line in Figure~\ref{fig:cmd_L22}. In the nearby regions where the \Gaia{} RVS is expected to be complete (regardless of extinction) we consider $n_{\rm L22}$ to be an accurate estimate of the true local number density up to the spectral bias of the MAFs and incompatibility between monochromatic and RVS band dust maps. Accordingly, we define $\epsilon_{\rm L22}$ as 
\begin{equation}
    \epsilon_{\rm L22}\equiv{n_{\rm obs}}/{n_{\rm L22}}
\end{equation}
and consider it to be an estimate for the true dust efficiency function within $d\sim2.5$~kpc.

The quoted uncertainties on $\epsilon_{\rm L22}$ include the reported error model for the extinctions in the L22 dust map. Additionally, we propagate measurement and statistical uncertainties by applying the techniques described in Appendix~\ref{app:uncertainties} to the extinction-corrected dataset. The L22 dust map model is extremely precise, and accordingly we find that the propagated errors in $\epsilon_{\rm L22}$ from the L22 dust map are completely subdominant to the other sources of error.

\section{Estimating Uncertainties}\label{app:uncertainties}
\begin{sidewaysfigure*}
    \centering
    \includegraphics[width=\textwidth]{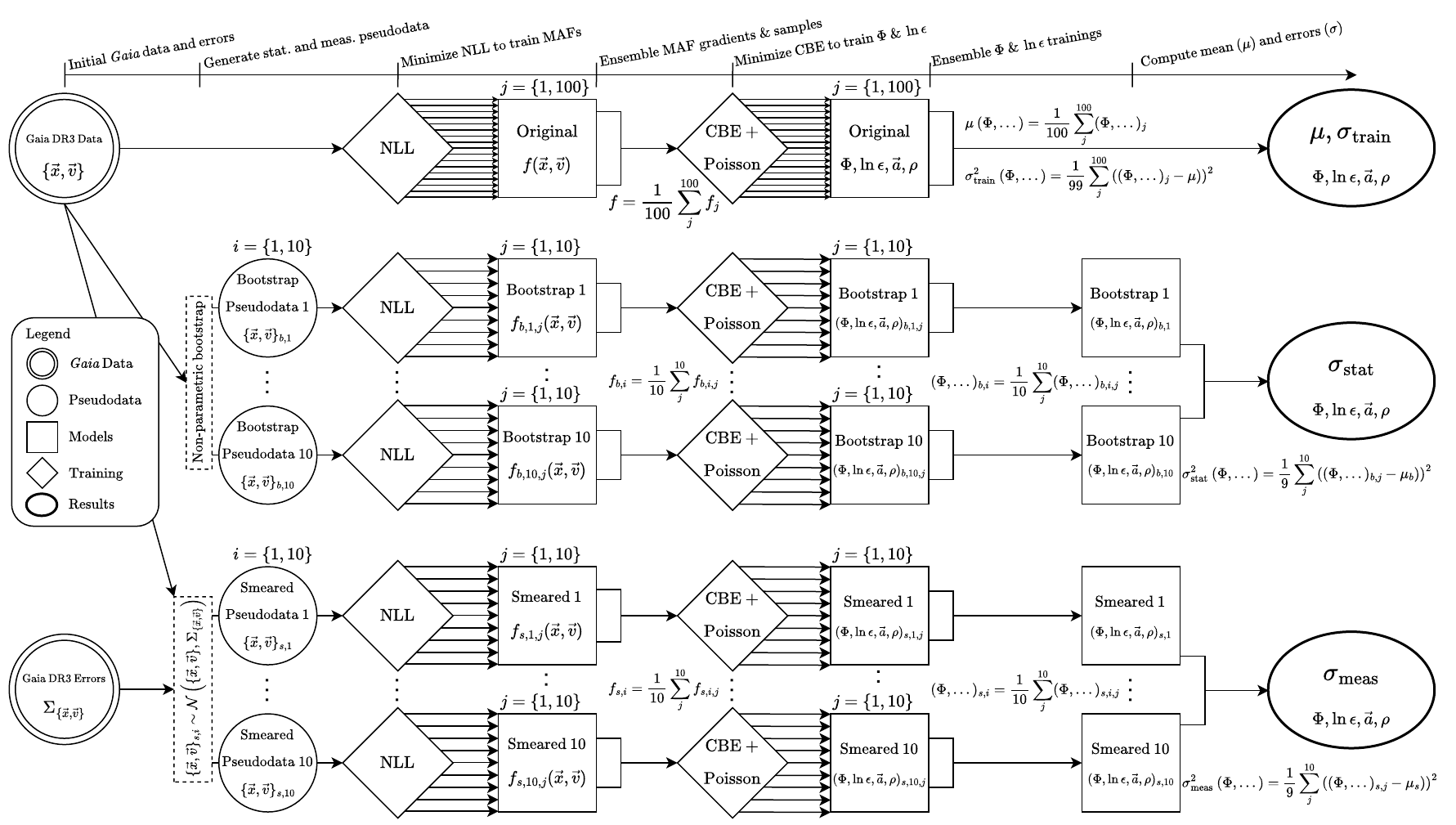}
    \caption{A schematic of the analysis pipeline presented in this work, from the \Gaia{} data to the dust efficiency $\epsilon$ and potential $\Phi$. From left: Input \Gaia{} RC/RGB dataset and the associated kinematic errors, which are used to create the 10 error-smeared and bootstrap-resampled datasets. An ensemble of flows are trained on each dataset, the gradients and samples of which are used to solve the CBE Eq.~\eqref{eq:CBE_eps} for $\epsilon(\vec{x})$ and $\Phi(\vec{x})$, and by extension $\vec{a}(\vec{x})$ and then $\rho(\vec{x})$ through the Poisson equation.
    Training uncertainties are estimated from the 100 training variations of $\epsilon$ and $\Phi$ on the original dataset, whereas measurement and statistical uncertainties are estimated from the variation of $\epsilon$ and $\Phi$ inferred across the different realizations of the 10 error-smeared and bootstrap-sampled datasets, respectively.}
    \label{fig:uncertainty_flowchart}
\end{sidewaysfigure*}

In order to characterize the training, measurement, and statistical uncertainties for our models of $f_{\rm obs}$, $\Phi$, and $\epsilon$ we follow a similar procedure as in Ref.~\cite{lim2023mapping}. In summary, we train many different realizations of these models across different iterations of the \Gaia{} dataset (plus errors) and network initializations. We then use the variation within the resulting $f_{\rm obs}$, $\Phi$, and $\epsilon$ networks as estimators of the various sources of error. A schematic of these error estimation techniques is shown in Figure~\ref{fig:uncertainty_flowchart}.

The primary models for $\Phi$ and $\epsilon$ presented in this work are each an ensemble average of $100$ networks with different initializations. The variation across these models serves as an estimate for the training uncertainty. When solving the CBE, we also sample and compute gradients from a separate ensemble of $100$ flows trained on the dataset introduced in Section~\ref{sec:dataset}, hereafter referred to as the ``original" dataset. Because the terms in the equilibrium CBE must be finely canceled to obtain $\partial f/\partial t \approx0$, it is especially important to use the ensemble average of a large number of flows to minimize noise in the computed the phase space gradients.

To incorporate the resulting uncertainties associated with the kinematic measurement errors, we generate ten realizations of the original dataset with each star perturbed by a randomly sampled value from Gaussians whose covariance matrices are set by the \Gaia{} DR3 kinematic error model. To marginalize over training variation when studying the measurement error, we train an ensemble of another ten normalizing flows on each noised realization of the dataset, and likewise use the samples and averaged gradients of this ensemble of flows when solving the CBE. Finally, we train ten copies of $\Phi$ and $\epsilon$ using each realization of the measurement error-smeared data to once again marginalize over training variation, and then define the measurement uncertainty as the variation across the ensembled models.

Similarly, we estimate statistical uncertainties associated with finite sample statistics by training $\Phi$ and $\epsilon$ using $f_{\rm obs}$ regressed to ten resampled realizations of the original dataset generated via the non-parametric bootstrap. Note that before the measurement error or statistical variance is computed, the models for $\Phi$ and $\epsilon$ are ensemble-averaged across ten different network initializations to minimize training variance.

To summarize, there are in total 100 models for $\Phi$ and $\epsilon$ based on the ``original" realization of the data, 100 models based on the ten ``smeared" realizations of the data (each ensembled over ten trainings), and 100 models based on the ten bootstrap-resampled realizations of the data (also ensembled over ten trainings). Although this comprehensive error analysis requires training 300 separate $f_{\rm obs}$, $\Phi$, and $\epsilon$ models, the latter two are relatively lightweight and quick to train compared to the MAF models that precede them. This approach provides robust estimates of a wide variety of sources of uncertainty in our analysis.

\bibliographystyle{JHEP}
\bibliography{refs}

\end{document}